%% file: site_version_franklin.tex
\documentclass[preprint,12pt]{elsarticle}

\usepackage{amssymb}

\journal{Applied Mathematical Modelling}

\begin{document}

\begin{frontmatter}



\title{LINEAR AND NONLINEAR INSTABILITIES OF A GRANULAR BED: DETERMINATION OF THE SCALES OF RIPPLES AND DUNES IN RIVERS \tnoteref{label_note_copyright} \tnoteref{label_note_doi}}

\tnotetext[label_note_copyright]{\copyright 2016. This manuscript version is made available under the CC-BY-NC-ND 4.0 license http://creativecommons.org/licenses/by-nc-nd/4.0/}

\tnotetext[label_note_doi]{Accepted Manuscript for Applied Mathematical Modelling, v. 36, p. 1057-1067, 2012, http://dx.doi.org/10.1016/j.apm.2011.07.058}


\author{Erick de Moraes Franklin}

\address{Instituto de Engenharia Mec\^anica - Universidade Federal de Itajub\'a\\
e-mail: franklin@fem.unicamp.br\\
Av. BPS, 1303 - Itajub\'a - MG - CEP: 37500-903\\
Brazil}

\begin{abstract}
Granular media are frequently found in nature and in industry and their transport by a fluid flow is of great importance to human activities. One case of particular interest is the transport of sand in open-channel and river flows. In many instances, the shear stresses exerted by the fluid flow are bounded to certain limits and some grains are entrained as bed-load: a mobile layer which stays in contact with the fixed part of the granular bed. Under these conditions, an initially flat granular bed may be unstable, generating ripples and dunes such as those observed on the bed of rivers. In free-surface water flows, dunes are bedforms that scale with the flow depth, while ripples do not scale with it. This article presents a model for the formation of ripples and dunes based on the proposition that ripples are primary linear instabilities and that dunes are secondary instabilities formed from the competition between the coalescence of ripples and free surface effects. Although simple, the model is able to explain the growth of ripples, their saturation (not explained in previous models) and the evolution from ripples to dunes, presenting a complete picture for the formation of dunes. 

\end{abstract}


\end{frontmatter}


\include{inst}






\bibliography{franklin}
\bibliographystyle{elsart-num}







\end{document}

%% file: inst.tex
\section{INTRODUCTION}

Granular media are frequently found in nature and in industry and their transport by a fluid flow is of great importance to human activities. Some examples are the aeolian transport of dust, the pneumatic conveyance of powder in the food industry, the conveyance of sand in hydrocarbon pipelines and the migration of sand dunes on both Earth and Mars. A better knowledge of the granular transport and of the associated instabilities is then of importance to understand nature and to improve industrial processes.

In some cases, the shear stresses exerted by the fluid flow are bounded to certain limits: the fluid flow can displace some grains, but not as a suspension. Instead, some grains are entrained by rolling or by executing small jumps, forming a mobile layer which stays in contact with the fixed part of the granular bed. If the fluid is a liquid, the thickness of this mobile layer is a few grain diameters \cite{Bagnold_1, Raudkivi_1}. This kind of transport is called bed-load.

A dimensionless parameter pertinent to the bed-load can be obtained from the dimensional analysis. This parameter, called Shields number $\theta$, is the ratio of the entraining force, of hydrodynamic nature, to the resisting force, related to the grains weight

\begin{equation}
\theta = \frac{\tau}{(\rho_{s}-\rho)gd}
\label{shields}
\end{equation}

\noindent where $\tau$ is the shear stress caused by the fluid on the granular bed, $d$ is the mean grain diameter, $\rho$ is the density of the fluid, $\rho_{s}$ is the density of the grains and $g$ is the gravitational acceleration. In order to obtain Eq.(\ref{shields}), the scale of the hydrodynamic force was considered as $\tau d^{2}$, and that of the resisting force as $(\rho_{s} - \rho)gd^{3}$. Bed-load takes place for $0.01\,\lesssim\,\theta\,\lesssim\,1$.

In the presence of bed-load, an initially flat granular bed may become unstable and give rise to bedforms. These forms, initially two-dimensional, may grow and generate patterns such as ripples or dunes.  In free-surface water flows, dunes are bedforms that scale with the flow depth, while ripples do not scale with it.

Of particular interest are the ripples and dunes appearing on the bed of rivers. The aquatic ripples create a supplementary friction between the bed and the water, affecting then the water depth and being related to flood problems. In the case of dunes, water flows can experiment local depth variations, affecting then navigation \cite{Engelund_Fredsoe}.

A balance between the local erosion and deposition of grains determines the conditions for the stability of a granular bed. Stability occurs whenever there is erosion at the crests of the granular bed, so that the amplitude of initial undulations decreases. The opposite situation means that the bed is unstable. Neutral stability exists if there is neither erosion nor deposition at the crests. The mass conservation of grains implies that there is erosion in regions where the gradient of the flow rate of grains is positive and deposition where it is negative. The phase lag between the flow rate of grains and the bedforms is then a stability criterion: if the maximum of the flow rate of grains is upstream the crests the bed is unstable, otherwise the bed is stable. Considering this, one manner to explain the bed stability is to understand the mechanisms creating a phase lag between the flow rate of grains and the shape of the granular bed.

These mechanisms were explained in a recent article \cite{Franklin_4}, in the specific case of granular beds sheared by turbulent boundary-layers of liquids, without free surface effects. Franklin (2010) \cite{Franklin_4} proposed that the mechanisms are essentially three: (i) the fluid flow perturbation by the shape of the bed, which is known to be the unstable mechanism \cite{Jackson_Hunt, Hunt_1, Weng}; (ii) the relaxation effects related to the transport of grains; and (iii) the gravity effects. Both $(ii)$ and $(iii)$ are stable mechanisms \cite{Valance_Langlois, Charru_3, Franklin_4}. Franklin (2010) \cite{Franklin_4} performed a linear stability analysis and showed that the length-scale of the initial bedforms varies with both the fluid flow conditions and the grains diameter.

A nonlinear stability analysis in the same scope of \cite{Franklin_4} was presented by \cite{Franklin_5}. The employed approach was the weakly nonlinear analysis \cite{Landau_Lifshitz, Schmid_Henningson, Drazin_Reid, Livre_Charru}, useful whenever a dominant mode can be proved to exist in the beginning of the nonlinear phase of the instability. This means that the modes resonating with the dominant one will grow much faster than the others, which can be neglected, so that the analysis is then made on a bounded number of modes. It was showed by \cite{Franklin_5} that, after the initial exponential growth (linear phase), the instabilities saturate, i.e., they attenuate their growth rate and maintain the same wavelength.

In a recent article, \cite{Fourriere_1} proposed that ripples are primary linear instabilities, while dunes are secondary instabilities formed from the coalescence of ripples. This was argued based on linear stability analyses for granular beds under turbulent streams, with and without free surface effects. Their results showed that the growth rate of the small wavelength modes (ripples) is four to six orders of magnitude greater than that for the large wavelength modes (dunes), so that the time scale to the linear development of dunes is many times greater than the time scales related to the formation and evolution of ripples. They concluded that dunes are formed by the coalescence of ripples. The formation of dunes by the coalescence of ripples was already proposed by \cite{Raudkivi_Witte, Raudkivi_2}.

Based on the hydrodynamics of a free-surface stream over a wavy bed, \cite{Fourriere_1} showed that the size of dunes is determined by the effects of the free surface. Dunes grow initially in subcritical regime and, as their sizes become comparable to the flow depth, they induce surface standing waves $180^o$ out-of-phase with respect to the dunes. As dunes continue to grow, their size will eventually affect locally the fluid flow regime, which will reach the transition from subcritical to supercritical. The latter is characterized by surface standing waves in phase with the dunes. In between the two regimes, there should be surface waves shifted of about $90^o$ downstream the dunes, which implies stability. This transition region, called by  \cite{Fourriere_1} \textit{resonance region}, determines the length-scale of dunes.

This paper presents a simplified analysis of the instabilities of a granular bed submitted to free-surface turbulent flows. The main purpose is to understand the growing and evolution of ripples and dunes observed in river and open-channel flows. The fundamental idea advanced by \cite{Fourriere_1} is employed here: ripples are formed as a primary linear instability while dunes are formed by the coalescence of ripples. A complete model able to explain the growth of ripples, their saturation and their evolution to dunes is then proposed, presenting a complete picture for the formation of dunes. In particular, the saturation of ripples was not explained in previous models for the formation of dunes and it is included for the first time here. In the present model, the formation of ripples is given by the analysis of \cite{Franklin_4}, and their nonlinear evolution is given by a modified version of \cite{Franklin_5}. The formation of dunes in then proposed by a simple analysis where the coalescence of ripples is the unstable mechanism while the effects of the free surface constitute the stable mechanism. 

The next section presents a brief summary of the results of the linear stability analysis of \cite{Franklin_4} and the following section proposes a modification on the nonlinear analysis of \cite{Franklin_5}. They are followed by a section discussing the free surface effects on the stability of the bed and by a section discussing the long-time evolution of bedforms. A conclusion section follows.

\section{LINEAR STABILITY ANALYSIS}
\label{section:linear}

In the last decades many linear stability analyses have been done in an attempt to understand the growth of bedforms on granular beds sheared by fluid flows. As examples, we may cite \cite{Kennedy, Reynolds, Engelund_1, Fredsoe_1, Engelund_Fredsoe, Richards, Elbelrhiti, Claudin_Andreotti, Fourriere_1}, among others. Usually, this kind of analysis find a most unstable mode, so that a length-scale and a celerity can be determined for this mode. However, given the linear character of the solution, the analyses are valid only during the initial phase of the instabilities, i.e., within a time-scale equal to $\sigma^{-1}$, where $\sigma$ is the growth rate of the most unstable mode.

Some of these works considered free-surface flows and found an unstable mode that scales with the flow depth. This mode, called dune mode, was shown in a recent work \cite{Fourriere_1} to have a much larger time-scale than another mode not scaling with the flow depth, called ripple mode. This discussion is left to section \ref{free_surface}.

A linear stability analysis on a granular bed sheared by a turbulent liquid flow, without free surface effects, was presented by \cite{Franklin_4}. The stability analysis was based on four equations, describing: (i) the fluid flow perturbation by the shape of the bed (in which the gravity effects were included), (ii) the transport of granular matter by a fluid flow, (iii) the relaxation effects related to the transport of grains, and (iv) the mass conservation of granular matter.

Due to the small character of the perturbations, the solutions were searched as plane waves and expressions for the growth rate $\sigma$ and the celerity $c$ of bedforms were found. The stability analysis showed the existence of a long-wave instability, with the fluid flow conditions, the relaxation effects and the gravity effects playing important roles. Based on these results, expressions for the the most unstable (or amplified) mode were determined as corresponding to $\partial\sigma / \partial k =0$, where $k$ is the wave-number. The expressions obtained by \cite{Franklin_4} for the wavelength $\lambda$, the growth rate $\sigma$ and the celerity $c$ of the most unstable mode are given below

\begin{equation}
\lambda_{max}\,\approx\,\frac{3B_A}{2B_e}L_{sat}
\label{L_max}
\end{equation}

\begin{equation}
\sigma_{max}\,\approx\,\frac{2}{9}\frac{B^3}{B_A^2}(B_A-2)Q_{sat}\frac{1}{(L_{sat})^2}
\label{s_max}
\end{equation}

\begin{equation}
c_{max}\,\approx\,\frac{B}{B_A}Q_{sat}\frac{1}{L_{sat}}
\label{c_max}
\end{equation}

\noindent where the subscript \textit{max} is related to the most unstable mode, $Q_{sat}$ is the saturated volumetric flow rate of grains by unit of width over a flat surface (basic state), $L_{sat}=\frac{u_*}{U_s}d$ is a distance called ``saturation length'' ($U_s$ is the typical settling velocity of one grain) and $A$, $B$, $B_e$ and $B_A$ are constants. Please refer to \cite{Franklin_4} for more details about the linear analysis.

Because this stability analysis didn't take into consideration the free surface, it is only applicable to the ripple mode. The most relevant result of this analysis concerns the dependence of the wavelength of the most unstable mode on the fluid flow, scaling as $\lambda_{max}\sim u_*$, where $u_*$ is the shear velocity. In a turbulent boundary-layer, the shear velocity is defined as $\tau=\rho u_*^2$. So, different from previous stability analyses for flows in turbulent regime without free surface, \cite{Franklin_4} proposed that the initial wavelength varies with the flow conditions if the carrier fluid is a liquid. This could explain some previous experimental results \cite{Kuru, Franklin_3}.

\section{NONLINEAR ANALYSIS}
\label{section:nonlinear}

In order to understand the evolution of ripples after the linear phase of the instability, i.e., in a time-scale greater than $\sigma^{-1}$, \cite{Franklin_5} presented a nonlinear stability analysis in the same scope of \cite{Franklin_4}. The approach used was the weakly nonlinear analysis \cite{Landau_Lifshitz, Schmid_Henningson, Drazin_Reid, Livre_Charru}, useful whenever a dominant mode can be proved to exist. This means that the modes resonating with the dominant one will grow much faster than the others, which can be neglected. The analysis is then made on a bounded number of modes. Charru (2007) \cite{Livre_Charru} presents a remarkable description of the weakly nonlinear approach, and of its application to the case of liquid films. The same ideas where applied by \cite{Franklin_5} in the case of a granular bed sheared by a turbulent flow of a liquid. The nonlinear analysis of \cite{Franklin_5} is briefly discussed next, and a small modification is proposed here.

The equations employed by \cite{Franklin_5} were the same as that of \cite{Franklin_4}. These equations were combined and a single equation could be obtained,

\begin{equation}
\partial_th+B_1(h)^2+B_2(\partial_xh)^2+B_3h\partial_xh+B_4h+B_5\partial_xh-ch=0
\label{nonlinear1}
\end{equation}

\noindent where $c$ is the celerity of the bedforms and $B_1$ to $B_5$ involve $Q_{sat}$, $L_{sat}$, $L_c$ and some constants, so that $B_1$ to $B_5$ are only functions of $u_*$ and $d$ ($L_c$ is a characteristic length of the bedform). They may then be treated as constants in an analysis of a given granular bed submitted to a given fluid flow.

In the nonlinear analysis of \cite{Franklin_5} the term $ch$ was treated as a constant, but this simplification seems inadequate. A modification is proposed here and the present analysis differs from that of \cite{Franklin_5} in its linear part. This difference is subtle, but it is of great importance because the nonlinear phase of the instability presents a fundamental mode that comes from the linear part.

In the weakly nonlinear analysis we are interested in the early stages of the nonlinear instability, when the exponential growth is no longer valid and the nonlinear terms become pertinent. In this case, it is a reasonable assumption to consider that the celerity $c$ of bedforms is approximately the celerity of the linear phase. The linear analysis of \cite{Franklin_4} showed that the celerity of the initial instabilities is given by $c\,\sim\,Q_{sat}/L{sat}$, so that $c\,\sim\,u_*(u_*d/U_s)^{-1}$. With this assumption, $c$ may be treated as a constant in an analysis of a given granular bed submitted to a given fluid flow. The last term of Eq. (\ref{nonlinear1}) is then considered here as varying with $h$.

In this problem the large scales are limited by periodicity and the small ones by the discrete nature of the grains, so that \cite{Franklin_5} considered a limited number of Fourier components as the normal modes. These modes were then inserted in Eq. (\ref{nonlinear1}) and the nonlinear terms were maintained. Normalizing the problem by its characteristic length ($k^{-1}$), and inserting the normal modes into Eq. (\ref{nonlinear1}), one can find the following equation 

\begin{eqnarray}
&\frac{1}{2}\displaystyle\sum_{n=-\infty}^\infty \left[\frac{dA_n}{dt}+A_n(B_4-c)+iB_5nA_n\right] e^{inx}  \nonumber \\
& + \frac{1}{2}\displaystyle\sum_{n=-\infty}^\infty \left[A_n^2B_1+B_2(inA_n)^2\right] e^{2inx} \nonumber \\
& + \frac{1}{2}\displaystyle\sum_{p=-\infty}^\infty \displaystyle\sum_{q=-\infty}^\infty \left[B_3A_piqA_q\right] e^{i(q+p)x}=0
\label{nonlinear2}
\end{eqnarray}

\noindent The difference between Eq.(\ref{nonlinear2}) and the corresponding equation in \cite{Franklin_5} lies in the first term on the left-hand side of Eq.(\ref{nonlinear2}).

Following the original idea of \cite{Landau_Lifshitz}, \cite{Franklin_5} neglected some terms in Eq.(\ref{nonlinear2}) because some other terms are expected to resonate with the linear part. By inspecting Eq.(\ref{nonlinear2}), \cite{Franklin_5} proposed that the third term may resonate with the linear part (in the first term) if $q+p=n$. Proceeding as \cite{Franklin_5} and keeping only the resonant terms of Eq.(\ref{nonlinear2}), we find

\begin{equation}
\frac{dA_n}{dt}\,=\,\sigma_nA_n\,+\,iB_3\displaystyle\sum_{p=-\infty}^\infty\left[pA_{p+n}A_p^*\right]
\label{nonlinear4}
\end{equation}
 
\noindent where $\sigma_n\,=\,c-B_4-inB_5$ (different from \cite{Franklin_5}).

Inspecting Eq.(\ref{nonlinear4}), it can be seen that $\sigma_n$ contains the linear part of the problem and that the non-linearities are in the third term. If the latter is neglected, which can be done in the initial phase of the instability, we find that the solution is stable for $\sigma_n<0$ and unstable for $\sigma_n>0$. In this case, $c-B_4$ is the parameter controlling the threshold of the instability: giving the scales of $c$, the basic state is stable when $u_*/L_{sat}\,<\,B_4$ and it is unstable when  $u_*/L_{sat}\,>\,B_4$, and the constant $B_4$ may be seen as a threshold value. We retrieve the conclusion of the linear analysis of \cite{Franklin_4}, that there is a cut-off wavelength that scales with both $u_*$ and $L_{sat}$, the small wavelengths being stable. It is important to note that this link with the linear analysis could not be done in \cite{Franklin_5} because, different from here, the term $ch$ was then treated as a constant.

Franklin (2010) \cite{Franklin_5} wrote Eq.(\ref{nonlinear4}) for the first three modes (for $n>0$) and, on the onset of the instability, showed that: (i) the second and higher modes can be written as functions of the first mode; (ii) the first one is a fundamental mode. An equation for the fundamental mode similar to the Landau Equation \cite{Landau_Lifshitz, Drazin_Reid} was then obtained

\begin{equation}
\frac{dA_1}{dt}\,=\,\sigma_1A_1\,-\,\kappa_LA_1\left| A_1\right|^2\,+\,O(A_1^5)
\label{nonlinear8}
\end{equation}

\noindent where $\kappa_L=-B_3^2/\sigma_2\,>\,0$.

From Eq.(\ref{nonlinear8}), we can see that the nonlinear term will saturate the instability in a time-scale greater than $\sigma^{-1}$: after the initial exponential growth, the instability will be attenuated and eventually reach a finite value for the amplitude, maintaining the same wavelength. This corresponds to a supercritical bifurcation \cite{Glendinning, Livre_Charru}. This result is corroborated by some experimental works \cite{Kuru, Franklin_3}, as discussed in \cite{Franklin_5}.

\section{FREE SURFACE EFFECTS: THE SCALES OF RIVER DUNES}
\label{free_surface}

It is known from measurements in the field that river beds have at least two characteristic wavelengths \cite{Engelund_Fredsoe, Fourriere_1}, one scaling with the grains diameter and with an inner layer close to the bed (but not with the flow depth), and another one scaling with the water depth. They correspond, respectively, to ripples and dunes.

In the two previous sections, the effects related to the presence of a free surface were not considered. This is valid for flows without free surface, or whenever the bedforms are much smaller than the flow depth, such as the ripples in river and open-channel flows. However, if we are interested in river bedforms, the presence of the free surface must be taken into account when considering dunes. River dunes are usually defined as bedforms whose size is of the same order of magnitude as the flow depth, so that the free surface affects the fluid flow near the bed.

In the last decades, many stability analyses have been done in order to understand the wavelength and the celerity of dunes in rivers \cite{Kennedy, Reynolds, Engelund_1, Richards, Engelund_Fredsoe, Coleman_2}. The great majority of these works found that the wavelength of the initial instabilities scales with the liquid depth, i.e., there would be a primary instability affected by the free surface. This primary instability must be strong enough to allow the growth of dunes in a time-scale smaller than, or of the same order as, that of ripples.

Richards \cite{Richards} performed a linear stability analysis that showed the presence of two unstable modes: one scaling with the liquid depth (dune mode) and the other scaling with the grains diameter, but independent of the free surface (ripple mode). He proposed that both modes would emerge as primary instabilities.

Coleman and Fenton (2000) \cite{Coleman_2} presented a reassessment of linear stability analyses describing river-bed instabilities based on the potential flow theory. They showed that potential flow analyses fail to predict the initial instabilities of an alluvial bed. They proposed, however, the existence of a resonance mechanism between surface and bed waves which can determine the length-scale of dunes developed from initial bedforms.

Raudkivi (2006) \cite{Raudkivi_2} presented a description of the generation of ripples and of their transition to dunes in alluvial beds. The author argued that the bed features are not oscillatory waves and that stability models are not adequate to explain the formation of ripples. Their generation is then described in terms of the turbulent structures of the fluid flow and the wavelength of ripples are proposed as a function of solely the grains diameter. However, this argument misses generality: (a) it cannot explain the formation of ripples in laminar flows; (b) the proposed equations relate the wavelength of ripples to a power of the grain diameter, so that the equations constants have dimensions (indicating a dimensional inconsistency). In the case of dunes, \cite{Raudkivi_2} proposed that they are formed by the coalescence of ripples, which agrees with previous experimental data presented in \cite{Raudkivi_2}.

Recently, \cite{Fourriere_1} presented a linear stability analysis for bedforms in rivers, in turbulent regime. In their analysis, they found that the ripple mode (not scaling with the liquid depth) has a growth rate many times greater than the dune mode (scaling with the liquid depth). Also, they found a range of wavelengths, scaling with the liquid depth, where the growth rate is strongly negative. They called this range ``resonance region''. Any bedform in this range has a strong negative growth rate.

Fourri\`ere et al. \cite{Fourriere_1} showed that the time scale for the growth of dunes as a primary linear instability is many times greater than that for the appearance of dunes by the coalescence of ripples (which have a much faster evolution). As a consequence, the dunes observed in rivers are the product of a secondary instability resulting from the coalescence of ripples, agreeing with \cite{Raudkivi_2, Coleman_2}. These dunes grow and reach a length-scale that corresponds to the resonance range, when their growth is then stopped by free surface effects. They concluded that ripples appear as a primary linear instability that eventually saturates (but they did not prove the saturation), and that dunes appear as a secondary instability resulting from the coalescence of ripples (unstable mechanism) and the free surface effects (stable mechanism in the resonance range).

This section is devoted to the study of bedforms observed in free-surface water flows, such as in open-channels and rivers. We know from previous works that in such cases there are at least two characteristic wavelengths, one scaling with the grains diameter and with an inner layer close to the bed (ripple mode), and another scaling with the fluid depth (dune mode). The ripple mode is unaffected by the free surface, so that the stability analyses summarized in sections \ref{section:linear} and \ref{section:nonlinear} are valid for this mode: the initial instabilities and their saturation can be predicted by these analyses. The dune mode, as proposed by the recent work of \cite{Fourriere_1}, is considered here as a secondary instability resulting from the competition between the coalescence of ripples and the free surface effects.

In order to obtain a complete picture for the formation of dunes from an initially flat granular bed, the models summarized in sections \ref{section:linear} and \ref{section:nonlinear} are complemented by an analysis of the effects caused by the free surface. Following \cite{Fourriere_1}, the fluid flow over dunes can be an unstable or stable mechanism relatively strong, i.e., the stability is determined by the fluid flow. To obtain the length-scale of dunes, it is proposed here a simple analytic manner to determine the shear stress caused by the flow on the dune surface. If the shear stress in out-of-phase with respect to the bedform, stability or instability may then be concluded.

The perturbation of the flow by a low hill can be determined by asymptotic methods. In the case of turbulent boundary-layers over hills of small aspect ratio, a two-region structure can be employed to determine the perturbed flow (a remarkable review is presented by \cite{Belcher_Hunt}). Furthermore, each of these two-regions are sometimes sub-divided in two layers in order to correctly match each other and the boundary conditions. Asymptotic methods have been formally applied to hills with aspect ratios of $h_{max}/(4L) < 0.05$ \cite{Jackson_Hunt, Sykes, Hunt_1} (the total length of the bedform is $\approx 4L$ and $h_{max}$ corresponds to the maximum value of the local height of the bedform $h$), but \cite{Carruthers_Hunt} showed that reasonable results are obtained when applied to slopes up to $h_{max}/(4L) = 0.3$ (dunes have aspects ratios of $h_{max}/(4L)=O(0.1)$).  In particular, when $h_{max}/(4L) = O(0.1)$, the obtained equations shall be applied to an envelope formed by the bedform and the recirculation bubble \cite{Weng}.

\begin{figure}
  \begin{center}
    \includegraphics[width=0.85\columnwidth]{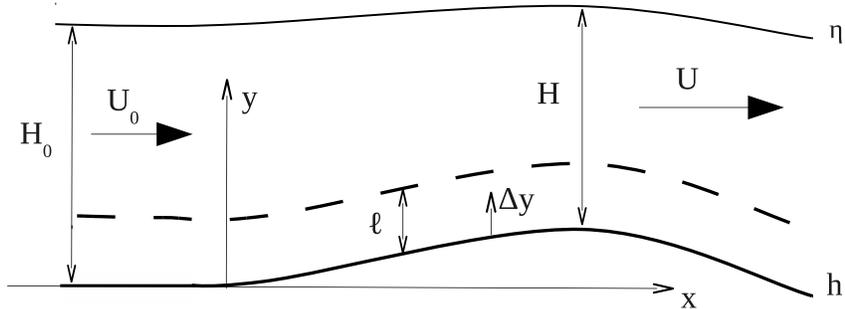}
    \caption{Definition of the free surface $\eta$, the bedform height $h$, the liquid depth $H$, the fluid velocity $U$ in the outer region, the height of the inner region $\ell$, the vertical coordinate employed in the outer region $y$ and the displaced vertical coordinate employed in the inner region $\Delta y$. The subscript $0$ corresponds to the fluid flow over a flat bed.}
    \label{fig:free_surf}
  \end{center}
\end{figure}

Following the same idea, the shear stress on the dune surface is determined here by an asymptotic approximation. The two-dimensional flow field is divided in two regions: an \textit{outer region} where the flow is far from the bed and is approximately potential; and an \textit{inner region}, smaller than the outer one and close to the bed, where shear stresses and viscous effects are strongly present. These regions are shown in Fig. \ref{fig:free_surf}.

The inner region is close enough to the bed, in a region where the turbulent vortices can adapt to equilibrium conditions with the mean flow over the dune. As explained by \cite{Belcher_Hunt}, the time scale for the dissipation of the energy-containing eddies is much smaller than the time scale for their advection, so that this region is in local-equilibrium. The local-equilibrium condition allows the use in this region of turbulent stress models, such as the mixing-length model. Also, as this region has a small thickness and follows the bedform without significant changes in thickness, the perturbations are driven by the pressure field of the outer region.

The outer region is far enough from the bed so that the energy containing vortices cannot adapt to equilibrium conditions with the mean flow. In other words, the time scale for the dissipation of the energy-containing eddies is much larger than the time scale for their advection, so that the flow is not in local-equilibrium \cite{Belcher_Hunt}. For this reason, the mean flow in this region is almost unaffected by the shear stress perturbations and a potential solution is expected at the leading order.

The division of the flow in only two regions, employed in this work, is a simple approach that allows a rough description of the flow over a bedform. However, in order to perfectly match the two regions, an intermediate layer (or region) must exist. In some works, each region is sub-divided in layers. For instance, \cite{Hunt_1} divided the inner region in two layers: one, closer to the bed surface, where the flow is determined by pressure and shear effects (the inertial effects are negligible) and which lower part match the boundary conditions on the bed; the other, closer to the outer region, where the flow is determined by pressure, shear and inertial effects and which upper part match the outer region. Hunt et al. (1988) \cite{Hunt_1} also divided the outer region in two layers. In the upper part, the ratio between the Reynolds stress gradient and the inertial effects is very small and the flow is approximately potential. In this layer the flow is dominated by pressure effects. The lower part must match the inner region, so that the flow in this layer is inviscid, but rotational.

\subsection{Outer region}
In the outer region, the perturbations induced by the bedform occur in a relative short length, so that the turbulence cannot adapt to the mean flow and the Reynolds stress perturbations are negligible \cite{Belcher_Hunt}. Then, a potential solution is expected at the leading order.

As this region is relatively far from the bed, the curvature of the streamlines is much smaller than near the bed, and the employed vertical coordinate is $y$, shown in Fig. \ref{fig:free_surf}. Also, due to the small curvature, we approximate the flow in this region as being uniform and perturbed in a potential manner in the first order on the small parameter $\varepsilon$. The flow is also affected by shear stress perturbations in an order smaller than, or equal to, $\varepsilon$ (this will be shown next).

Given the small aspect ratios (the aspect ratios reported for river dunes are $O(0.1)$), a reasonable choice for the small parameter is $\varepsilon\,=\,h/H_0$. Expansions of the mean velocity of the fluid flow $U$ and of the free surface position $\eta$ in power series of the small parameter $\varepsilon$ give

\begin{equation}
U\,=\,U_A\,+\,\varepsilon U_B\,+\,O(\varepsilon^2)\,+\,O(\varepsilon_1)
\label{exp_U}
\end{equation}

\begin{equation}
\eta\,=\,\eta_A\,+\,\varepsilon\eta_B\,+\,O(\varepsilon^2)\,+\,O(\varepsilon_1)
\label{exp_n}
\end{equation}

\noindent where $U_A$, $U_B$, $\eta_A$, and $\eta_B$ are the expansion terms and are $O(1)$. $\varepsilon_1$ is the small parameter to be used in the inner region. The aforementioned approximation of an uniform basic flow means that $U_A=0$ is a solution.

If a potential flow approximation is expected in the first order in $\varepsilon$, then the Bernoulli equation, together with the mass conservation, may be employed in the outer region (Eqs. \ref{bernoulli1} and \ref{mass1} , respectively)

\begin{equation}
\frac{U^ 2}{2}+g\eta\,=\,\frac{U_0^ 2}{2}
\label{bernoulli1}
\end{equation}

\begin{equation}
U\left( H_0-h+\eta\right)\,=\,U_0H_0
\label{mass1}
\end{equation}

\noindent where $H$ is the flow depth and the subscript $0$ corresponds to the fluid flow over a flat bed.

By inserting the expansions (Eqs. \ref{exp_U} and \ref{exp_n}) in Eqs. (\ref{mass1}) and (\ref{bernoulli1}), and separating the terms by their respective orders, one obtains

\begin{equation}
O(1) : \, \begin{array}{c}U_A+U_A\frac{\eta_A}{H_0}\,=\,U_0 \\ \, \\ \frac{1}{2}U_A^2+g\eta_A\,=\,\frac{1}{2}U_0^2 \\ \end{array}
\label{O_1}
\end{equation}

\noindent where $U_A=U_0$ and $\eta_A=0$ is a solution (and it is physically expected), and

\begin{equation}
O(\varepsilon) : \, \begin{array}{c}-U_A+U_A\frac{\eta_B}{H_0}+U_B+\frac{U_B}{H_0}\eta_A\,=\,U_0 \\ \, \\ U_AU_B+g\eta_B\,=\,0 \\ \end{array}
\label{O_e}
\end{equation}

The solution of the system given by Eq. (\ref{O_e}) is $\eta_B=H_0(Fr^2/(Fr^2-1))$ and $U_B=-U_0(1/(Fr^2-1))$, where

\begin{equation}
Fr\,=\,\frac{U_0}{\sqrt{gH_0}}
\end{equation}

\noindent is the Froude number: the ratio between the velocity of the mean flow and the celerity of surface gravity waves. The expansions given by Eqs. (\ref{exp_U}) and (\ref{exp_n}) then become

\begin{equation}
U\,=\,U_0\left[1-\frac{h}{H_0}\left(\frac{1}{Fr^2-1}\right)\right]\,+\,O(\varepsilon^2)\,+\,O(\varepsilon_1)
\label{eq:U}
\end{equation}

\begin{equation}
\eta\,=h\left(\frac{Fr^2}{Fr^2-1}\right)\,+\,O(\varepsilon^2)\,+\,O(\varepsilon_1)
\label{eq:n}
\end{equation}

An expansion for the pressure in the outer region can be written based on Eq. (\ref{eq:n}). In non-dimensional form, the pressure $P_{outer}^{ad}$ is

\begin{equation}
P_{outer}^{ad}\,=\,\frac{H}{H_0}\,=1\,+\,\frac{h}{H_0}\left(\frac{1}{Fr^2-1}\right)\,+\,O(\varepsilon^2)\,+\,O(\varepsilon_1)
\label{eq:Pad}
\end{equation}

\subsection{Inner region}
In this region the flow is in local equilibrium, i.e., as the flow pass the dune the eddies adapt to be in equilibrium with the mean flow, in a time scale much shorter than the one to advect them \cite{Belcher_Hunt}. Because this region is relatively close to the bed, the curvature of the streamlines approximately follows the bedform. In order to better describe the flow in this region, it is preferable to employ the displaced vertical coordinate $\Delta y$ shown in Fig. \ref{fig:free_surf}

\begin{equation}
\Delta y\,=\,y-h
\end{equation}

Solutions for the inner region of turbulent boundary-layers perturbed by hills have been found by some authors, as for instance \cite{Jackson_Hunt, Sykes, Hunt_1, Weng}. In all cases, the mixing-length model was employed as a closure for the turbulent stresses and the nature of the perturbation was the same: the flow perturbations in the inner region were considered as driven, in the first order in $\varepsilon_1$, by the pressure field of the outer region.

These solutions for the inner region are expected to be similar to the present problem, given that all cases concern local-equilibrium conditions. The form of the solution presented by \cite{Weng} is employed here, but with the pressure perturbation found previously for the outer region. In the inner region, as proposed by \cite{Jackson_Hunt,Hunt_1, Weng}, the flow over the bedform can be found as expansions in power series of the small parameter $\varepsilon_1=ln\left(\ell /z_0\right)$, where $\ell$ is the thickness of the inner region and $z_0$ is the roughness height. For turbulent flows over sand grains, $\varepsilon_1$ is $O(0.1)$, so that is has the same order as $\varepsilon$. From a balance between the acceleration terms and the stress terms in the longitudinal momentum equation, and employing the mixing-length model, \cite{Jackson_Hunt} found that the thickness $\ell$ of the inner region is given by

\begin{equation}
\frac{\ell}{L}ln\left( \frac{\ell}{z_0}\right)\,=\,2\kappa^2
\end{equation}

\noindent where $\kappa=0.41$ is the K\'arm\'an constant.

\subsection{Matching and shear stress on the bed}
Because the stability of a granular bed depends on the bed-load flow rate and, in their turn, the formulae for the bed-load flow rate are functions of the shear stress on the bed, we are particularly interested in the latter. The perturbed (non-dimensional) shear stress on the bed ($\Delta y \rightarrow 0$) in the longitudinal direction $\tau_{pert}$ can be found by matching the inner and the outer regions.

In order to perfectly match the inner and the outer regions, an intermediate region between them needs to be taken into account. In this region, the flow can be considered as inviscid, but it is rotational. In an attempt to keep the model simple, we do not explicitly create an intermediate region. Instead we employ the expression for $\tau_{pert}$ obtained by \cite{Weng} who considered, in its deduction, the intermediate region 

\begin{equation}
\tilde{\tau}_{pert}=F\{P_{outer}^{ad}\}\frac{k^2}{|k|} \left[ 1+\frac{2lnL|k|+4\gamma +1+i\pi sign(k)}{ln\left(\ell /z_0\right) }\right]
\label{eq:weng}
\end{equation}

\noindent where the tilde denotes the Fourier space, $F$ is the Fourier transform operator, $\gamma\,=\,0.577216$ is the Euler's constant, $sign(k)$ is the sign of $k$ and $L$ is the half length (longitudinal distance between the crest and the position where the bedform height is half of its maximum value, so that the total length of the bedform is $\approx 4L$). 

Following the idea of \cite{Sauermann_2, Kroy_C, Kroy_A}, the terms that are functions of $ln\left(\ell /z_0\right)$ can be grouped together and, as their variation is very small, they may be considered as constants. This gives

\begin{equation}
\tilde{\tau}_{pert}=f(Fr)AF\{h\} \left[ k+iBk \right]
\label{eq:tau_fourier}
\end{equation}

\noindent where $A$ and $B$ are constants, and $f(Fr)$ is constant for a given fluid flow

\begin{equation}
f(Fr)=\frac{-1}{Fr^2-1}
\label{eq:C_1}
\end{equation}

\begin{equation}
A=\left[ 1+\frac{2ln(1/4)+4\gamma +1 }{ln\left(\ell /z_0\right) }\right]
\label{eq:A}
\end{equation}

\begin{equation}
B=\frac{1}{A}\frac{\pi}{ln\left(\ell /z_0\right)}
\label{eq:B}
\end{equation}

\noindent Variations of $ln\left(\ell /z_0\right)$ over three decades do not change the order of magnitude of $A$ and $B$. For example, for $O(10^3)<ln\left(\ell /z_0\right)<O(10^5)$, $A=O(1)$ and $B=O(0.1)$ (both having positive values). In the real space, Eq.(\ref{eq:tau_fourier}) becomes

\begin{equation}
\tau_{pert}=f(Fr)A\left(\frac{h}{H_0}\,+\,B\partial_{x}h\right)
\label{eq:tau_real}
\end{equation}

\noindent and the stress on the bed surface is then

\begin{equation}
\tau\,=\,\tau_0\left( 1+\tau_{pert}\right)
\label{eq:total_stress}
\end{equation}

\noindent where $\tau_0$ is the shear stress on a flat bed (basic state).

For a particular bedform, Eq.(\ref{eq:tau_real}) gives the perturbation of the shear stress. A rapid examination in Eq.(\ref{eq:tau_real}) informs about the phase shift between the shear stress perturbation and the bedform. Not considering $f(Fr)$ in Eq.(\ref{eq:tau_real}), the equation is composed of a symmetric term $A\frac{h}{H_0}$ and of an anti-symmetric term $AB\partial_{x}h$ with respect to the bedform. As $A=O(1)$ and $B=O(0.1)$, the combined effect of these two terms is to cause a small upstream phase shift of the the shear stress with respect to the bedform.

A stronger effect on the phase shift can be caused by the term $f(Fr)$. When $0<Fr<1$, $f(Fr)>0$ and there is no contribution of $f(Fr)$ to shift the phase. In this case, $\tau_{pert}$ is slightly shifted upstream the bedform, so that the fluid flow is an unstable mechanism. When $Fr>1$, $f(Fr)<0$ and $f(Fr)$ causes a $180^o$ phase shift. In this case, $\tau_{pert}$ is almost in phase opposition with respect to $h$, having a small upstream contribution (due to the anti-symmetric term), so that the fluid flow is a stable mechanism. When $Fr=1$, there is a singularity in $f(Fr)$, and $\tau_{pert}$ cannot be determined.

\begin{figure}
  \begin{center}
    \includegraphics[width=0.6\columnwidth]{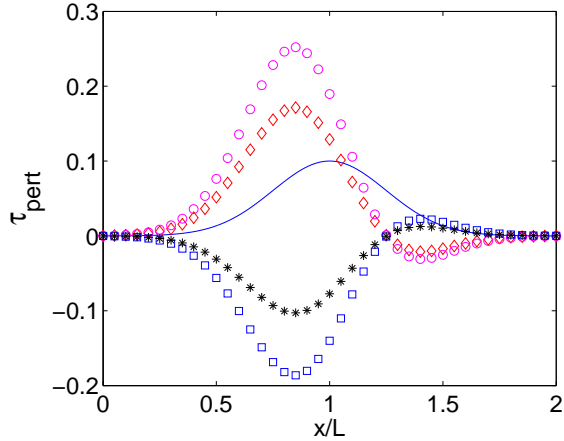}
    \caption{Longitudinal evolution of the perturbation of the shear stress $\tau_{pert}$, for different Froude Numbers $Fr$. The symbols $\Diamond$, $\bigcirc$, $\square$ and $\ast$ correspond to $Fr=0.5$, $Fr=0.7$, $Fr=1.3$ and $Fr=1.5$, respectively. The continuous line corresponds to the normalized bedform $h/H_0$.}
    \label{fig:pert_shear}
  \end{center}
\end{figure}

As an example, the perturbation of the shear stress $\tau_{pert}$ caused by a Gaussian bedform was computed employing Eq.(\ref{eq:tau_real}) for different Froude numbers $Fr$ and is shown in Fig. \ref{fig:pert_shear}. In this figure, the aspect ratio of the bedform is $h_{max}/(4L)\,=\,0.1$, and the normalized bedform $h/H_0$ is also shown. The stronger contribution to the phase shift is seen to be given by the flow regime.

\begin{figure}
  \begin{center}
    \includegraphics[width=0.6\columnwidth]{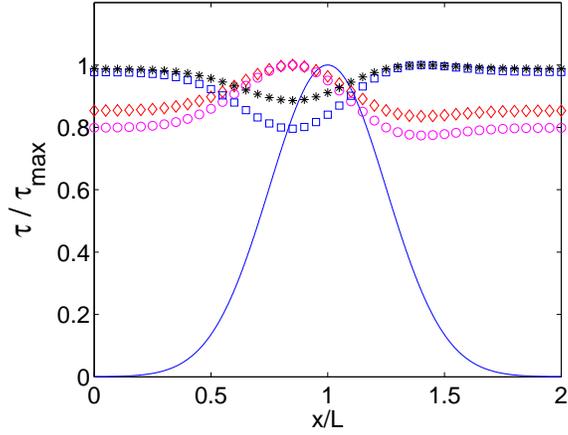}
    \caption{Longitudinal evolution of the normalized shear stress $\tau/\tau_{max}$, for different Froude Numbers $Fr$, where $\tau_{max}$ is the maximum of the shear stress. The symbols $\Diamond$, $\bigcirc$, $\square$ and $\ast$ correspond to $Fr=0.5$, $Fr=0.7$, $Fr=1.3$ and $Fr=1.5$, respectively. The continuous line corresponds to a normalized bedform $h/h_{max}$, where $h_{max}$ is the maximum height of the bedform.}
    \label{fig:shear}
  \end{center}
\end{figure}

The shear stress $\tau$ may be computed by Eq.(\ref{eq:total_stress}). Figure \ref{fig:shear} presents the shear stress $\tau$ on the bed surface for the same bedform of Fig. \ref{fig:pert_shear}, for different Froude Numbers $Fr$. The shear stress on a flat bed was evaluated as $\tau_0=\rho u_{*,0}^2$, where $u_{*,0}$ is the shear velocity over a flat bed. The shear velocity $u_{*,0}$ was computed by employing friction factor correlations for two-dimensional open-channel flows \cite{Yen}.

\begin{figure}
  \begin{center}
    \includegraphics[width=0.6\columnwidth]{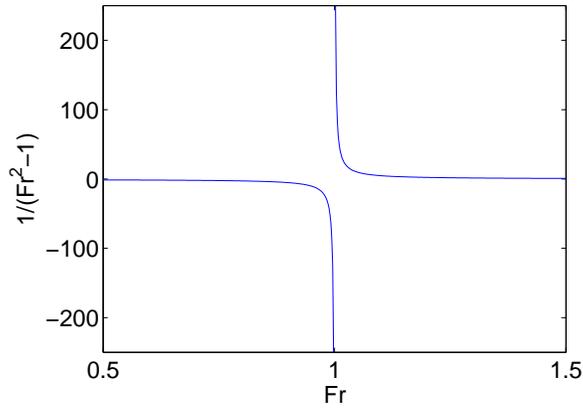}
    \caption{$f(Fr)$ in the vicinity of the singularity $Fr=1$. The ordinate was limited to the range $-250$ to $+250$.}
    \label{fig:f_Fr}
  \end{center}
\end{figure}

At $Fr=1$, the term $f(Fr)$ in Eq.(\ref{eq:tau_real}) presents a singularity. At this singularity and in its vicinity, as shown in Fig. \ref{fig:f_Fr}, it is impossible to predict $\tau_{pert}$ due to the different limits of $f(Fr)$ ($-\infty$ or $+\infty$ depending on the direction the limit is taken). For this reason, the behavior at $Fr=1$ is not discussed here, instead we discuss the transition from $0<Fr<1$ to $Fr>1$. In the vicinity of $Fr=1$, between $0.9<Fr<1.1$, the strong variations of $f(Fr)$ give abnormal strong amplitudes for the shear stress perturbation.

In subcritical regime $0<Fr<1$, the maximum of the shear stress on the bed occurs upstream the maximum of the bedform, so that the fluid flow will be always an unstable mechanism. This shift is due to the anti-symmetric term of Eq.(\ref{eq:tau_real}), which has a negligible variation with the flow conditions (Eq.\ref{eq:B}). As discussed briefly in the Introduction section, and in \cite{Franklin_4} in more detail, provided that the upstream phase shift of the fluid flow is larger than the downstream phase shift of the relaxation effects of the granular transport, the bed is unstable and bedforms tend to grow. For bedforms larger than the ripple length-scale, such as dunes, the phase shift of the fluid flow is the greater (if the fluid is a liquid) and bedforms will grow.

In supercritical regime $Fr>1$, the phase shift between the shear stress and the bedform is $180^o$ minus the shift caused by the the anti-symmetric term of Eq.(\ref{eq:tau_real}). In this case, the fluid flow will be always a stable mechanism. As the fluid flow is strongly shifted downstream, the bed is then stable.

\subsection{The celerity of dunes}
The celerity of dunes can be estimated in a simple manner by employing a simplified transport model for the grains together with the mass conservation applied to the granular bed (Eq. \ref{exner})

\begin{equation}
\partial_{t}h+\partial_{x}q=0
\label{exner}
\end{equation}

\noindent where $q$ is bed-load flow rate by unit of width. Considering that the shear stress on the bed is proportional to the square of the mean velocity (which is equivalent to employ a friction factor), and approximating the bed-load flow rate by Eq. (\ref{qsat})

\begin{equation}
\frac{q_{sat}}{Q_{sat}}=(1+\tau_{pert})^{3/2}
\label{qsat}
\end{equation}

\noindent where $q_{sat}$ is the saturated volumetric flow rate of grains by unit of width, we may write the bed-load flow rate (by unit of width) as $q\,\approx\,C_1U^3$, where $C_1$ is a constant. In this case

\begin{equation}
\frac{\partial h}{\partial t}\,+\,c_0\frac{\partial h}{\partial x}\,=\,0
\label{eq:wave}
\end{equation}

\noindent where

\begin{equation}
c_0\,\approx\,-\frac{3C_1U^2U_0}{H_0}\frac{1}{Fr^2-1}
\label{eq:c0}
\end{equation}

\noindent Equation (\ref{eq:wave}) is an advection equation for the bedform, with celerity $c_0$. When $c_0>0$, the bedforms interacting with the free surface are called dunes, and when $c_0<0$ these forms are called anti-dunes. \cite{Engelund_Fredsoe}.

\section{Discussion}

The free surface effects described in the previous section, together with the formation of ripples given by the linear analysis of \cite{Franklin_4} and with the evolution of ripples given by the nonlinear analysis of \cite{Franklin_5}, form a simple and complete picture for the formation of dunes in rivers.

Considering as initial condition a flat river bed, ripples will grow as a primary instability that can be predicted by \cite{Franklin_4}. The linear analysis of \cite{Franklin_4} predicts that the length of ripples scales as $\lambda_{max}\,\sim\,L_{sat}$ (the length of ripples can be computed by Eq.(\ref{L_max})), where $L_{sat}=\frac{u_*}{U_s}d$ is a distance called ``saturation length''. From the proposed scaling, besides its dependence on the grains diameter $d$, the length-scale of ripples depends also on the conditions of the water stream. Concerning their celerity, it also scales with $L_{sat}$, and can be predicted by Eq.(\ref{c_max}).

After their initial (and exponential) growth, the ripples have their growth rate attenuated and reach eventually a finite size. This can be predicted by the nonlinear analysis of \cite{Franklin_5}, that was modified in section \ref{section:nonlinear}.

The ripples are displaced by the fluid flow and, as their celerity is inversely proportional to their size \cite{Franklin_4}, they interact and coalescence may occur. The proposition of \cite{Fourriere_1} that the time scale for the coalescence of ripples is much smaller than the time scale for the direct formation of dunes as a primary instability is adopted here. This allows the determination of the scales of dunes by considering the stability effects caused by the free surface.

In this work a simple model was used to capture the free surface effects. The water stream over a dune was divided in two regions: an outer region where the fluid flow perturbations (by the bedform) are essentially potential; and an inner region where the turbulent shear stresses need to be taken into account. A simple solution based on the Bernoulli equation was found for the outer region, which was then matched with an inner solution adapted from \cite{Weng}. As the bed instability is directly related to the shear stress caused by the fluid flow, the obtained expressions for the perturbation concern the shear stress on the bed (Eqs.\ref{eq:tau_real} and \ref{eq:total_stress}). With regard to the celerity of dunes and anti-dunes, an advection equation (Eq.\ref{eq:wave}) was obtained by employing a simplified transport model for the grains together with the mass conservation applied to the granular bed.

The examination of Eq.(\ref{eq:tau_real}) at subcritical regime ($0<Fr<1$) shows that the bedforms are unstable, tending to increase in size. This is consistent with the potential analysis of the problem (which is equivalent to consider only the outer region). From Eqs.(\ref{eq:U}) and (\ref{eq:n}), $\eta<0$ and $U>U_0$, meaning that the free surface above the bedform is a trough: the surface gravity waves are $180^o$ out-of-phase with respect to the bedforms and the maximum of the fluid velocity is around the crest, with an upstream out-of-phase component which is only captured when the inner region is considered. Also, in this case $c_0>0$, so that the bedforms have a downstream celerity and are usually called dunes \cite{Engelund_Fredsoe}.

At supercritical regime ($Fr>1$), Eq.(\ref{eq:tau_real}) shows that the bedforms are stable, tending to decrease in size. This is also consistent with the potential analysis of the problem. From Eqs.(\ref{eq:U}) and (\ref{eq:n}), $\eta>0$ and $U<U_0$, meaning that the free surface above the bedform is a bump: the surface gravity waves are in phase with respect to the bedforms and the minimum of the fluid velocity is around the crest, with an out-of-phase component which is only captured when the inner region is considered. Also, in this case $c_0<0$, so that the bedforms have an upstream celerity and are usually called anti-dunes \cite{Engelund_Fredsoe}.

The analysis of the flow at subcritical and supercritical regimes shows that the bedforms are unstable when $0<Fr<1$ and stable when $Fr>1$, so that the size of the dunes is determined by the condition $Fr\approx 1$ and their wavelength scales with $H_0$. If we consider that $0.1H_0<h_{max}<H_0$, then the length-scale of dunes is $H_0<\lambda_{dunes}<10H_0$, which is in agreement with the observed river dunes \cite{Guy, Allen, Julien, Coleman_3}. The interaction between the surface waves and the bedforms bounds the length-scale of dunes: dunes grow in subcritical regime as a result of ripples coalescence until their size approaches the regime transition ($Fr=1$). Then, the water stream becomes a stable mechanism and limits the size of dunes.

In the vicinity of the transition from subcritical to supercritical, the solutions diverge due the singularity of Eq.(\ref{eq:tau_real}) (at $Fr=1$), and the strong amplitudes predicted for the shear stress perturbations are probably overestimated. This is a drawback of the model, principally because the anti-dunes are usually observed when $Fr\approx 1$. The present model predicts that the anti-dunes would decrease as soon as $Fr>1$, until $0<Fr<1$ is reached, becoming then dunes. However, the formation of anti-dunes is still an open question and they are in many instances the result of sudden changes in the water stream. Then, for steady-state water streams the present model can be used to describe the formation of dunes from the coalescence of ripples.

\section{CONCLUSION}

This paper presented a complete and simple model able to predict the formation and the evolution of sand patterns observed in rivers and in open-channels. In particular, the saturation of the amplitude of ripples, not explained in previous models, was included in the present one. Based on previous works \cite{Raudkivi_2, Franklin_4, Franklin_5, Fourriere_1}, it was argued here that ripples are a primary linear instability, which saturates with the same wavelength predicted by the linear analysis, and that dunes are a secondary instability. In order to obtain the length-scale of dunes, a simplified model was proposed in which the coalescence of saturated ripples is the unstable mechanism, while the fluid flow, influenced by the free surface, is the stable mechanism. The proposed model retains only the terms and equations essentials to understand the problem and is able to predict the length-scale and the celerity of both ripples and dunes.

%% file: site_version_franklin.bbl
\begin{thebibliography}{10}
\expandafter\ifx\csname url\endcsname\relax
  \def\url#1{\texttt{#1}}\fi
\expandafter\ifx\csname urlprefix\endcsname\relax\def\urlprefix{URL }\fi

\bibitem{Bagnold_1}
R.~A. Bagnold, The physics of blown sand and desert dunes, Chapman and Hall,
  1941.

\bibitem{Raudkivi_1}
A.~Raudkivi, Loose boundary hydraulics, 1st Edition, Pergamon Press, 1976.

\bibitem{Engelund_Fredsoe}
F.~Engelund, J.~Fredsoe, Sediment ripples and dunes, Ann. Rev. Fluid Mech. 14
  (1982) 13--37.

\bibitem{Franklin_4}
E.~Franklin, Initial instabilities of a granular bed sheared by a turbulent
  liquid flow: length-scale determination, J. Braz. Soc. Mech. Sci. Eng. 32~(4)
  (2010) 460--467.

\bibitem{Jackson_Hunt}
P.~S. Jackson, J.~C.~R. Hunt, Turbulent wind flow over a low hill, Quart. J. R.
  Met. Soc. 101 (1975) 929--955.

\bibitem{Hunt_1}
J.~C.~R. Hunt, S.~Leibovich, K.~Richards, Turbulent shear flows over low hills,
  Quart. J. R. Met. Soc. 114 (1988) 1435--1470.

\bibitem{Weng}
W.~S. Weng, J.~C.~R. Hunt, D.~J. Carruthers, A.~Warren, G.~F.~S. Wiggs,
  I.~Livingstone, I.~Castro, Air flow and sand transport over sand-dunes, Acta
  Mechanica (1991) 1--21.

\bibitem{Valance_Langlois}
A.~Valance, V.~Langlois, Ripple formation over a sand bed submitted to a
  laminar shear flow, Eur. Phys. J. B 43 (2005) 283--294.

\bibitem{Charru_3}
F.~Charru, Selection of the ripple length on a granular bed sheared by a liquid
  flow, Physics of Fluids 18~(121508).

\bibitem{Franklin_5}
E.~Franklin, Nonlinear instabilities on a granular bed sheared by a turbulent
  liquid flow, J. Braz. Soc. Mech. Sci. Eng. in press.

\bibitem{Landau_Lifshitz}
L.~Landau, E.~Lifshitz, Fluid mechanics, Pergamon Press, 1959.

\bibitem{Schmid_Henningson}
P.~Schmid, D.~Henningson, Stability and transition in shear flows, Springer,
  2001.

\bibitem{Drazin_Reid}
P.~G. Drazin, W.~R. Reid, Hydrodynamic Stability, 2nd Edition, Cambridge
  University press, 2004.

\bibitem{Livre_Charru}
F.~Charru, Instabilit\'es hydrodynamiques, 1st Edition, EDP Sciences, 2007.

\bibitem{Fourriere_1}
A.~Fourri\`ere, P.~Claudin, B.~Andreotti, Bedforms in a turbulent stream:
  formation of ripples by primary linear instabilities and of dunes by
  nonlinear pattern coarsening, J. Fluid Mech. 649 (2010) 287--328.

\bibitem{Raudkivi_Witte}
A.~Raudkivi, H.~Witte, Development of bed features, J. Hydraul. Eng. 116 (1990)
  1063--1079.

\bibitem{Raudkivi_2}
A.~Raudkivi, Transition from ripples to dunes, J. Hydraul. Eng. 132 (2006)
  1316--1320.

\bibitem{Kennedy}
J.~F. Kennedy, The mechanics of dunes and antidunes in erodible-bed channels,
  J. Fluid Mech. 16 (1963) 521--544.

\bibitem{Reynolds}
A.~J. Reynolds, Waves on the erodible bed of an open channel, J. Fluid Mech. 22
  (1965) 113--133.

\bibitem{Engelund_1}
F.~Engelund, Instability of erodible beds, J. Fluid Mech. 42 (1970) 225--244.

\bibitem{Fredsoe_1}
J.~Fredsoe, On the development of dunes in erodible channels, J. Fluid Mech. 64
  (1974) 1--16.

\bibitem{Richards}
K.~J. Richards, The formation of ripples and dunes on an erodible bed, J. Fluid
  Mech. 99 (1980) 597--618.

\bibitem{Elbelrhiti}
H.~Elbelrhiti, P.~Claudin, B.~Andreotti, Field evidence for
  surface-wave-induced instability of sand dunes, Nature 437~(04058).

\bibitem{Claudin_Andreotti}
P.~Claudin, B.~Andreotti, A scaling law for aeolian dunes on mars, venus,
  earth, and for subaqueous ripples, Earth Plan. Sci. Lett. 252 (2006) 20--44.

\bibitem{Kuru}
W.~C. Kuru, D.~T. Leighton, M.~J. McCready, Formation of waves on a horizontal
  erodible bed of particles, Int. J. Multiphase Flow 21~(6) (1995) 1123--1140.

\bibitem{Franklin_3}
E.~Franklin, Dynamique de dunes isol\'ees dans un \'ecoulement cisaill\'e,
  Ph.D. thesis, Universit\'e de Toulouse (2008).

\bibitem{Glendinning}
P.~Glendinning, Stability, instability and chaos: an introduction to the theory
  of nonlinear differential equations, Cambridge University Press, 1999.

\bibitem{Coleman_2}
S.~Coleman, J.~Fenton, Potential-flow instability theory and alluvial stream
  bed forms, J. Fluid Mech. 418 (2000) 101--117.

\bibitem{Belcher_Hunt}
S.~Belcher, J.~C.~R. Hunt, Turbulent flow over hills and waves, Ann. Rev. Fluid
  Mech. 30 (1998) 507--538.

\bibitem{Sykes}
R.~Sykes, An asymptotic theory of incompressible turbulent boundary-layer flow
  over a small hump, J. Fluid Mech. 101 (1980) 647--670.

\bibitem{Carruthers_Hunt}
D.~Carruthers, J.~Hunt, Fluid mechanics of airflows over hills: turbulence,
  fluxes, and waves in the boundary layer, Atmospheric Processes over Complex
  Terrain 23 (1990) 83--108.

\bibitem{Sauermann_2}
G.~Sauermann, Modeling of wind blown sand and desert dunes, Ph.D. thesis,
  Universit\"at Stuttgart (2001).

\bibitem{Kroy_C}
K.~Kroy, G.~Sauermann, H.~J. Herrmann, Minimal model for sand dunes, Phys. Rev.
  Lett. 88~(054301).

\bibitem{Kroy_A}
K.~Kroy, G.~Sauermann, H.~J. Herrmann, Minimal model for aeolian sand dunes,
  Phys. Rev. E 66~(031302).

\bibitem{Yen}
B.~Yen, Open channel flow resistance, J. Hydraul. Eng. 128~(1) (2002) 20--39.

\bibitem{Guy}
H.~Guy, D.~Simons, E.~Richardson, Summary of alluvial channel data from flume
  experiments, U.S. Geol. Survey Prof. Paper 462-I (1966) 1--96.

\bibitem{Allen}
J.~Allen, Physical processes of sedimentation, American Elsevier, 1970.

\bibitem{Julien}
P.~Julien, G.~Klaassen, Sand-dune geometry of large rivers during floods, J.
  Hydraul. Eng. 121~(9) (1995) 657--663.

\bibitem{Coleman_3}
S.~Coleman, V.~Nikora, S.~McLean, T.~Clunie, T.~Schlicke, B.~Melville,
  Equilibrium hydrodynamics concept for developing dunes, Physics of Fluids
  18~(105104).

\end{thebibliography}
